# Optomechanical THz detection with a sub-wavelength resonator

**Cherif Belacel, Yanko Todorov, Stefano Barbieri, Djamal Gacemi, Ivan Favero, and Carlo Sirtori**

*Laboratoire Matériaux et Phénomènes Quantiques, Université Paris Diderot, Sorbonne Paris Cité, CNRS-UMR 7162, 10 rue Alice Domont et Léonie Duquet, 75013 Paris, France*

**The terahertz spectral domain offers a myriad of applications spanning chemical spectroscopy, medicine, security and imaging[1]. It has also recently become a playground for fundamental studies of light-matter interactions[2-6]. Terahertz science and technology could benefit from optomechanical approaches, which harness the interaction of light with miniature mechanical resonators[7,8]. So far, optomechanics has mostly focused on the optical and microwave domains, leading to new types of quantum experiments[9-11] and to the development of optical-microwave converters[12-14]. Here we demonstrate an integrated meta-atom[15] terahertz resonator with a flexible part acting as a mechanical oscillator. In this device free space terahertz photons are collected by the resonator and induce high frequency currents and charges that, in turn, couple to the mechanical degrees of freedom. The resulting mechanical motion is read-out optically, allowing our device to function as a compact and efficient terahertz detector at room temperature. Furthermore the device operates at high modulation frequencies (>10MHz), well beyond the cut-off frequencies of Golay cells, pyroelectric detectors and cryogenic semiconductor bolometers[16,17]. Notably, our experiments unambiguously reveal an instantaneous terahertz detection mechanism arising from a nano-scale Coulomb interaction, with a Noise Equivalent Power (NEP) that is potentially frequency independent. Alongside this effect, our compact geometry allows for an uncooled bolometric terahertz detection[18] with extremely short heat diffusion times (few microseconds) and high detectivity. The optomechanical nanodevice introduced in this work is suitable for integration on local probes or within imaging arrays, opening new prospects for sensing in the terahertz domain, as well as for applications relying on high-speed, bright terahertz sources such as quantum cascade lasers[19,20] and synchrotrons[21].**

The device realized in this work is shown in Figure 1 and consists of an asymmetric split ring resonator (SRR), obtained by depositing a metal pattern on a GaAs/AlGaAs layered semiconductor structure (Methods). By using standard semiconductor etching technology, the narrow arm of the resonator has been processed into a cantilever with a high aspect ratio (length $L$=15.7 μm, width $w$=444 nm, and thickness $t$=470 nm, including 320 nm of GaAs and 150 nm of gold layers). As shown in the blown-up part of Fig. 1, the free end of the cantilever forms an arm of 308 nm wide capacitive gap ($d_{gap}$) of the SRR. The geometrical dimensions of the SRR have been chosen to set its fundamental electromagnetic resonance in the Tera Hertz (THz) domain. When the structure is resonantly excited by the incident THz radiation, a dynamic distribution of charges with opposite signs appears on both sides of the gap. This results into a quasi-static Coulomb force that attracts the cantilever towards the opposite end of the gap, setting it in motion. The amplitude of this mechanical oscillation is directly related to the intensity of the incident THz wave, and, as explained further below, can be read out optically with high sensitivity.

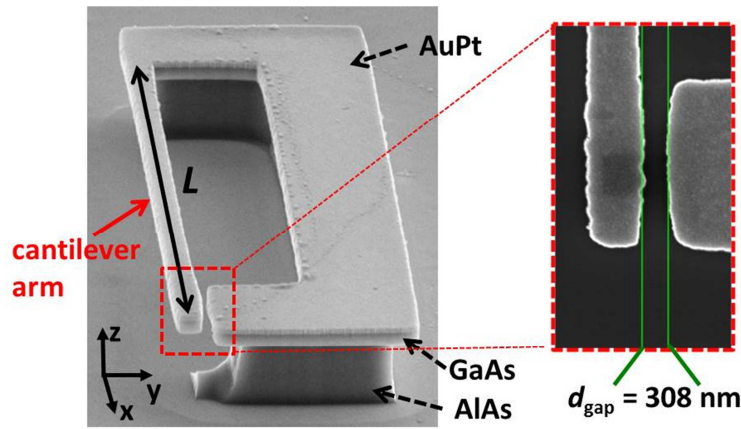

**Figure 1: THz optomechanical device.** Side view of the SRR resonator obtained with a scanning electron microscope. The SRR is closed on one side by a thin cantilever arm. Different metallic and semiconductor layers of the structure are visible, as indicated. The blown up image shows the capacitive gap between the cantilever arm and the rest of the SRR.

In Figure 2, we report on the THz electromagnetic resonance of our structure. Figure 2a displays the SRR THz response measured in transmission spectroscopy with a Fourier Transform Interferometer and a cooled Ge bolometer (see Methods). In order to increase the amplitude of the transmission feature we used a dense array of nominally identical SRRs[22,23] as shown in the inset. The SRR resonance appears in the spectrum of Fig. 2a as a Lorentzian dip with a central frequency $\omega_{THz}/2\pi$=2.7 THz and a quality factor $Q_{THz}$=8.4. In Fig. 2b we plot the electric energy density provided by finite element method simulations. The electric energy is strongly localized in the SRR gap, favoring the Coulomb force acting on the cantilever. In Fig 2c, we also plot the Eddy currents induced in the SRR, which turn out to play an important role for the photo-thermal effects in the structure, as explained further.

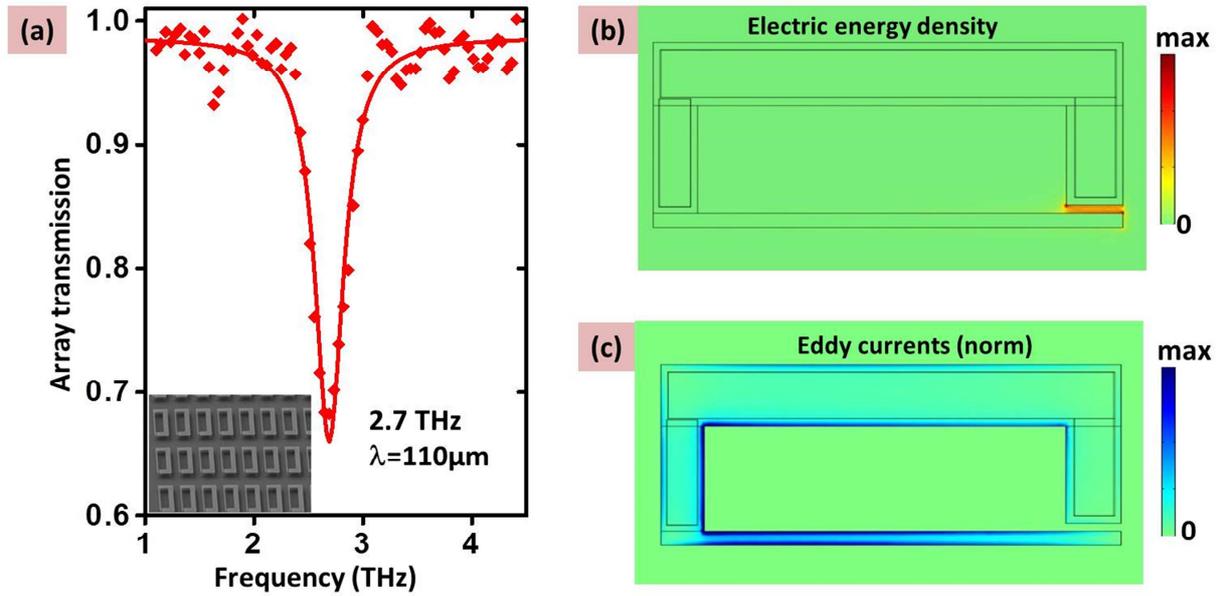

**Figure 2: Electromagnetic resonance of the SRR**. **a** Far infrared transmission spectrum of an array of identical SRRs obtained with a FTIR spectrometer and a Ge bolometer. The THz beam was focused on the sample with the help of 4 parabolic mirrors. Owing to the large spot-size of the focal point (~1mm²) we used a dense array of SRRs (see Methods). **b**, Distribution of the electric energy density in the SRR, obtained with a finite difference domain simulator. The electric energy is strongly localized in the gap, as expected. **c**, Illustration of the THz Eddy currents (norm) induced in the metal layers at resonance.

To characterize the mechanical modes of the cantilever, we focus a near infrared (NIR, $\lambda$=940 nm) laser beam at its end. The Brownian motion of the cantilever produces intensity fluctuations in the reflected beam that are recorded as the signal difference between two balanced Si photo-diodes connected to a spectrum analyzer (SA). In Fig. 3a we show the resulting Radio Frequency (RF) spectra of the fundamental in-plane ($\alpha_1$) and out-of-plane ($\beta_1$) flexural modes, illustrated by numerical simulations in Fig. 3b. The corresponding frequencies and quality factors are respectively $f_{\alpha 1}$=0.86 MHz, $Q_{\alpha 1}$=70; $f_{\beta 1}$=0.91 MHz, $Q_{\beta 1}$=94. At frequencies around 5 MHz, second order resonances, noted as $\alpha_2$ and $\beta_2$ are also visible (Figs. 3c and 3d). In all cases, the Brownian motion is well fitted by the analytical expression of the noise spectral power density $S_{yy}(f)$ from a dumped mechanical oscillator model[8], taking into account the detection noise floor. The knowledge of the cantilever dimensions and composition allows determining its effective mass $m_{eff}$=8.5 pg, and hence a known value for the room temperature peak noise spectral density $S_{yy}(f_m)=2k_B T Q_m / m_{eff}(2\pi f_m)^3$ for each resonance in the fits of Figs. 3a and 3c. This permits a precise calibration of the cantilever displacements measured with the present technique.

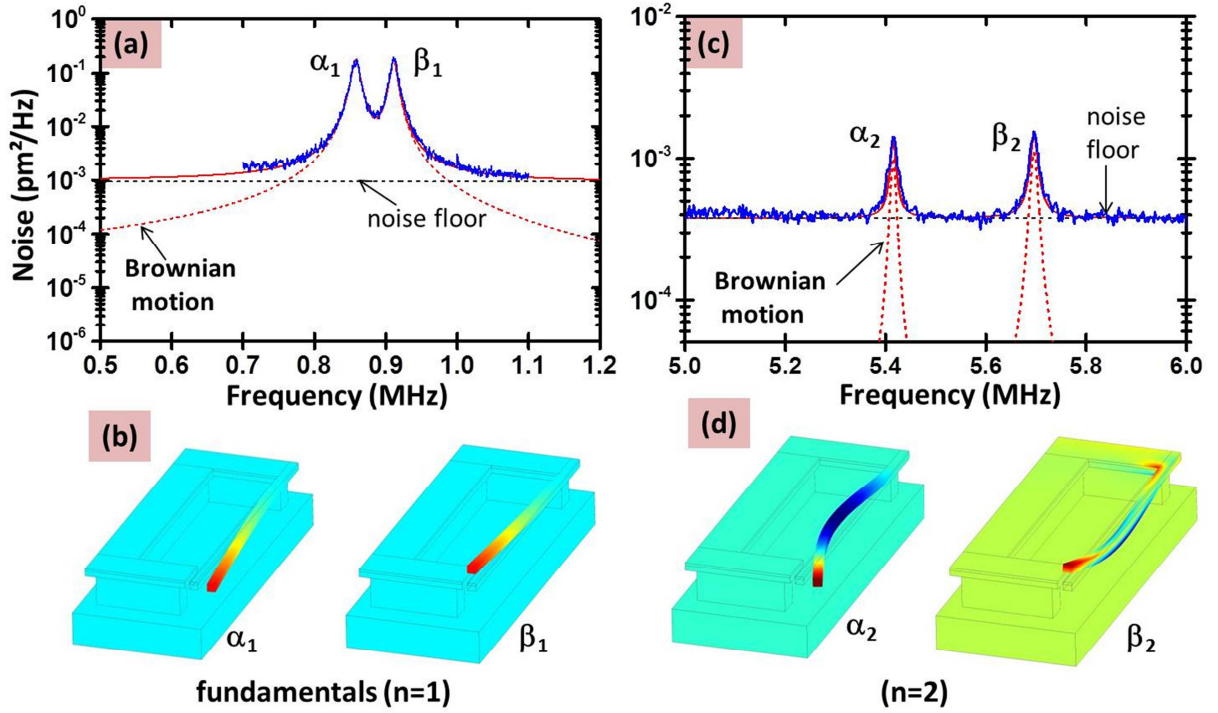

**Figure 3: Mechanical spectroscopy of the cantilever modes**. **a**, Fundamental resonances visible in the Brownian motion of the cantilever measured with a spectral analyzer and the experimental setup of Fig. 4 (a) (blue curve). The red curve is a fit resulting from the sum of the spectral noise of two harmonic oscillators (red dashed curve) and the noise floor of the signal from the balanced photo-diodes (black dashed curve). The two resonances, labelled $\alpha_1$ and $\beta_1$, correspond respectively to the motion of the cantilever in the plane of the SRR ($\alpha_1$) and perpendicular to it ($\beta_1$). **b,** Illustration of the modes $\alpha_1$ and $\beta_1$ obtained by numerical modelling. **c,** Spectra of the second order mechanical resonances, $\alpha_2$ and $\beta_2$, with the corresponding fit. **d,** Illustration of the cantilever bending for the second order mechanical modes. Note that the noise floor decreases with frequency, owing to a 1/f component.

The cantilever movement induced by the Coulomb attractive force will have a stronger effect on the in-plane (α) mechanical modes, and is described by the following equation of motion, derived from an effective capacitor on spring model (Methods):

(1) $$\frac{d^2 y}{dt^2} + \frac{\omega_{\alpha m}}{Q_{\alpha m}} \frac{dy}{dt} + \omega_{\alpha m}^2 y = -\frac{W_{THz}(t)}{m_{eff}} \frac{d\ln C_{eff}}{dy}\bigg|_{y=d_{gap}}$$

Here $y$ is the position of the cantilever end (Fig. 1), $C_{eff}(y)$ is the effective capacitance of the charge distribution of the THz mode, and $W_{THz}(t)$ is the time dependent electric energy stored in the SRR at resonance. Note that the electric energy oscillates in the THz range, $W_{THz}(t) \sim \cos^2(\omega_{THz} t)$, i.e. six orders of magnitude higher than the mechanical frequency $\omega_{\alpha m}$, a situation reminiscent of cavity optomechanics at optical frequencies[7]. As a result the cantilever is only sensitive to the value of the electric energy $<W_{THz}>$ averaged over a THz oscillation cycle. The latter can be expressed as $<W_{THz}> = P_{THz} Q_{THz}/2\omega_{THz}$, where $P_{THz}$ is the THz power dropped in the SSR. Then, according to Eq.(1), the net mechanical effect of a continuous THz wave impinging on the detector is to displace the rest position of the cantilever. This effect can be resonantly exalted if the incident THz intensity is modulated close to the cantilever mechanical frequency. For a sinusoidal modulation $P_{THz} = P^0_{THz}(1+\cos(\omega t))/2$ we can define a frequency-dependent internal responsivity of the system as the ratio between the amplitude of the resulting forced mechanical motion $y(\omega)$ and the peak THz power $P^0_{THz}$ coupled into the SRR:

$$(2) \quad R_{in}(\omega) = \frac{|y(\omega)|}{P^0_{THz}} = \frac{Q_{\alpha m}}{2 m_{eff} \omega^2_{\alpha m}} \frac{Q_{THz}}{d^{eff}_{gap} \omega_{THz}} |H_\alpha(\omega)|$$

Here $H_\alpha(\omega) = (\omega^2_{\alpha m}/Q_{\alpha m})(\omega^2_{\alpha m} - \omega^2 + i\omega \omega_{\alpha m}/Q_{\alpha m})$ is the transfer function of the harmonic oscillator, normalized such as $|H_\alpha(\omega=\omega_{\alpha m})|=1$. We introduced the effective capacitive gap $d^{eff}_{gap} = ((d\ln C_{eff}/dy)|_{y=d_{gap}})^{-1}$ which depends on the details of the charge distribution at resonance. For our structure, we estimate $d^{eff}_{gap} \sim$ 800 nm based on simulations of the electrostatic energy as a function of the cantilever endpoint displacement $y$ (Methods). Using Eq.(2) together with the value of the expression of the peak noise density of the Brownian motion, we evaluated the internal Noise Equivalent Power (NEP) for this mechanism, defined as $NEP=S_{yy}(f_m)^{1/2}/R_{in}$. For the current geometry we obtain a peak responsivity $R_{in}(\omega=\omega_{\alpha 1}) \sim 34$ fm/nW and $NEP \sim 16$ nW/Hz$^{1/2}$.

To compare the present device to other optomechanical systems[8], we have also evaluated the frequency pull parameter $g_{om} = \omega_{THz}/d^{eff}_{gap} =$ 21 GHz/nm and the amplitude of zero-point mechanical motion $y_{ZPF} = \sqrt{\hbar/2 m_{eff} \omega_{\alpha m}}$ = 33 fm, which yield a single-photon optomechanical coupling $g_0 = g_{om} y_{ZPF}$ = 0.7 MHz. This value is commensurable with the mechanical frequency of the cantilever, which, in the quantum regime, would imply the possibility to resolve individual THz photons by recording the induced quantized mechanical displacement. Such non-demolition ponderomotive probe of electromagnetic energy was already considered in the early work of Braginsky[24], and while it has remained out of reach for optical and microwave optomechanical settings, it could seemingly be accessible in the THz domain.

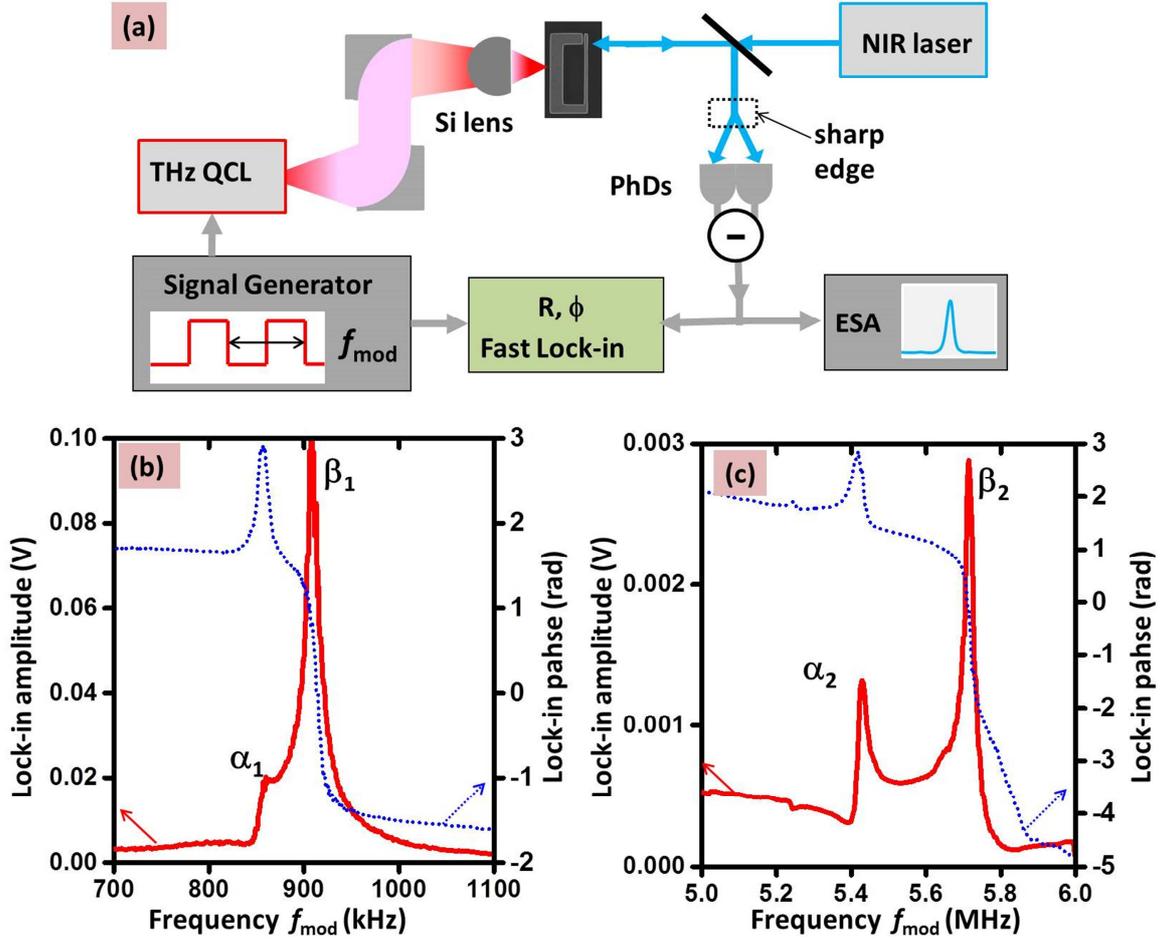

**Figure 4: THz optomechanical detection a,** Experimental set-up of the THz optomechanical detection scheme. The radiation emitted by the THz QCL is collected with two parabolic mirrors and focused on a single SRR with the help of a silicon hyperhemispherical lens. The QCL drive current is modulated at a frequency $f_{mod}$, close to the mechanical resonance of the cantilever. The cantilever movement is read-out optically as described in Methods. **b,** Amplitude (full curve) and phase (dotted curve) from the lock-in, obtained as the modulation frequency of the QCL is swept around the first pair of resonances $\alpha_1$ and $\beta_1$. In this experiment, the Brownian motion spectrum monitored by the SA is identical to Fig. 3a. **c,** Lock-in amplitude (full curve) and phase (dotted curve) as the frequency $f_{mod}$ is swept around the second order modes, with Brownian spectra identical to Fig. 3c.

The experimental optomechanical setup is reported in Fig. 4a. As a THz source, we use a quantum cascade laser (QCL)[25], with an emission frequency of 2.6 THz and a maximum calibrated emitted power of 4.7 mW. The THz radiation from the QCL is collected with two parabolic mirrors and focused on a single SRR with the help of a silicon hyperhemispherical lens[26], positioned on the backside of the GaAs substrate. The THz QCL is either driven in pulsed mode, or in continuous wave with its current modulated with a signal generator at the frequency $f_{mod}$. The THz laser beam thus excites the cantilever oscillations, which are read-out optically as described above.

The coherent mechanical response of the SRR induced by the THz radiation is studied by sweeping the QCL modulation frequency $f_{mod}$ and recording both amplitude and phase on the lock-in amplifier connected to the photodiodes. Figure 4b shows the response around the fundamental mechanical modes ($\alpha_1$ and $\beta_1$), while Fig. 4c corresponds to the second order modes, ($\alpha_2$ and $\beta_2$). As can be seen an important fraction of signal is collected not only from the in-plane modes $\alpha_1$ and $\alpha_2$ driven by the Coulomb force, but also from the out-of-plane modes $\beta_1$ and $\beta_2$, which in the present experimental configuration have larger response. The latter cannot be excited efficiently by the Coulomb force since the out-of-plane variation of the capacitance is negligible in the present geometry. Instead, the out-of-plane ($\beta$) modes are efficiently excited by a photo-thermal force arising from the bi-layer structure of the cantilever[18]. Indeed, in our device the incoming THz radiation generates heat through the Eddy currents resonantly excited in the SRR, as illustrated in Fig. 2c. Since the GaAs and Au layers have different thermal expansion coefficients[18,27], the temperature stress, also modulated at $f_{mod}$, leads to the resonant excitation of the $\beta$-modes. The photo-thermal force induced by the THz Eddy currents can be modelled by solving the dynamical heat-transfer equation in harmonic regime and by calculating the thermally induced stress, as outlined in the Methods. As the resulting z-displacement of the cantilever is proportional to the absorbed THz power, we ultimately obtain the photo-thermal responsivity:

$$(3) \quad R_{int}^{ph}(\omega) = \frac{|z(\omega)|}{P_{THz}^0} = \frac{f_{ph}^0 Q_{\beta m}}{m_{eff} \omega_{\beta m}^2} |Y(\omega) H_\beta(\omega)|, \quad Y(\omega) = \sum_{n=0}^{\infty} \frac{A_n}{i\omega\tau_0 + (2n+1)^2}$$

Here $f_{ph}^0$ is the amplitude of the photo-thermal force per absorbed unit THz power, which is a function of the bi-layer geometry and some physical constants (Methods), $n$ is an integer, $A_n$ is a series of dimensionless coefficients that depend on the spatial profiles of the Eddy currents, the temperature rise, and the shape of the cantilever mechanical mode (Methods). The quantity $H_\beta(\omega)$ is the transfer function of the harmonic oscillator $\beta$. Values and more details are provided in Methods. The time constant $\tau_0$ is expressed as $\tau_0 = 4\tau/\pi^2$, where $\tau = L^2/D$ is the thermal diffusion time (D is the diffusion coefficient provided in Methods), which in the present geometry is $\tau = 3$ μs. The function $Y(\omega)$ defined in Eq.(3) takes into account the retardation effects of heat propagation along the cantilever, and basically acts as a low pass frequency filter on the instantaneous harmonic oscillator response $H_\beta(\omega)$[27]. For the $\beta_1$ mode Eq. (3) provides a peak responsivity of 860 fm/nW, i.e. approximately 25 times larger than the response for the quasi-static (Coulomb) mechanism associated to the $\alpha_1$ mode $R_{in}(\omega)$, which explains quantitatively our experimental results (see Fig. 5b,5e and the discussion below).

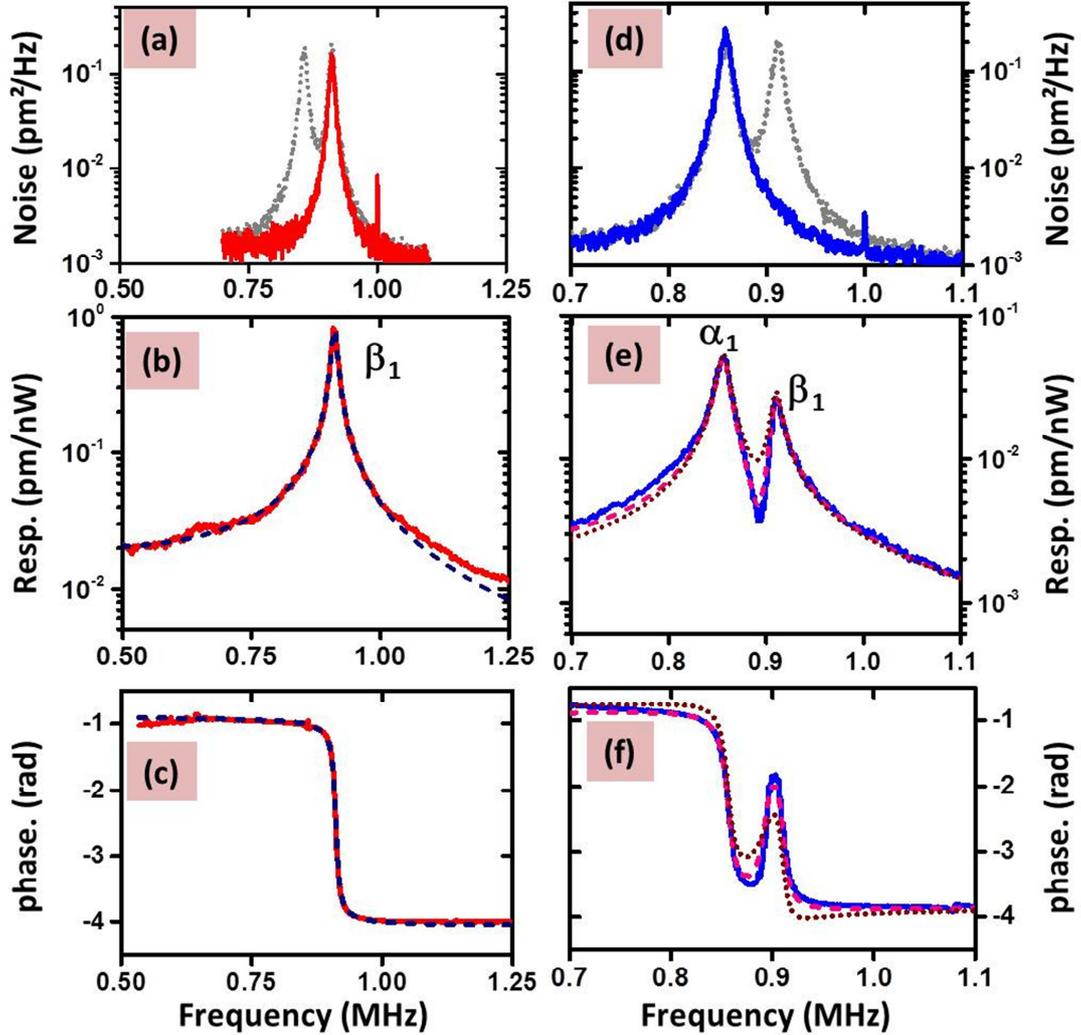

**Figure 5: Data and models. a,** Optimized NIR detection of the $\beta_1$ resonance (red curve). The dotted curve reproduces the spectrum from Fig. 3a. **b, c,** Lock-in amplifier amplitude (5b) and phase (5c) measured in this condition (continuous curve). The dashed curve sare fits are based on Eq.(3). **d,** Optimized NIR detection of the $\alpha_1$ resonance (blue curve). **e,f,** Measured amplitude (5e) and phase (5f) (continuous curves). Dotted curves: fit assuming that the force acting on the $\alpha_1$ resonance is described by Eq.(2), while the force acting on the $\beta_1$ resonance is described by Eq.(3). Dashed curves: fit including a photo-thermal contribution to the optomechanical force acting on $\alpha_1$.

A key feature of our NIR detection scheme is that the overall signal is a linear superposition of in-plane and out-of-plane modes, $|l_y y(t)+l_z z(t)|$, where the coefficients $l_{y,z}$ are real and can be adjusted by changing the position of the focal point of the NIR laser on the cantilever, as illustrated in Figure 5. This allows favoring the detection of in-plane or out-of-plane modes, in order to study the corresponding THz response mechanisms. The coefficients $l_y$ and $l_z$ are directly provided by fitting the RF Brownian spectra, as shown in Fig. 2 (Methods).

In the data presented in Fig. 5a,b,c, the detection of the $\beta_1$ out-of-plane mode has been favored in the Brownian spectra. The THz optomechanical response, measured with the lock-in amplifier, is shown in Fig. 5b (amplitude) and Fig. 5c (phase). Data are very well fitted by the model from Eq.(3) (dashed curves), assuming that only the $\beta_1$ mode is present, setting the absolute values of the peak responsivity and phase. The peak responsivity of the $\beta_1$ mode was quantified experimentally by using the calibration of the QCL output power (4.7mW) and the maximum mechanical displacement of the cantilever ($z_{max}$=71 nm), in the case were the QCL was modulated with a 16.5% duty-cycle square wave. This yielded an external responsivity of 50 fm/nW, which is compatible with the value 860 fm/nW estimated by our model when considering 5% coupling efficiency of the Si lens (Methods). The corresponding internal *NEP* is estimated at 0.4 nW/Hz$^{0.5}$, which is already close to the state of the art of Golay cells[16,17].

In contrast, for the data presented in Fig. 5d,e,f, the detection of the $\alpha_1$ in-plane mode is favored over $\beta_1$. Even though the $\beta_1$ resonance is no longer visible in the Brownian motion spectra (Fig. 5d), it still contributes significantly to the forced oscillations induced by the THz radiation, as shown in Fig. 5e. Furthermore, the phase shown in Fig. 5f displays a peculiar feature around the $\beta_1$ resonance. A first attempt to fit the data is done by assuming that only Coulomb instantaneous force acts on the in-plane mode $\alpha_1$ (Eq.(2)), while only a retarded photo-thermal force acts on the out-of-plane mode $\beta_1$ (Eq.(3)). This is shown by the dotted lines in Figs. 5e and 5f. Even though a good agreement is found, the amplitude of the phase variations in-between the two resonances is not reproduced exactly. A better fit (dashed curves in Figs. 5e and 5f) is instead obtained by adding to the instantaneous Coulomb force exciting the $\alpha_1$ mode an additional retarded photo-thermal contribution of the form $iC_{ph}H_\alpha(\omega)/(1+i\omega\tau)$, as described in the model of Ref. [27]. Here the constant $C_{ph}$ is of similar magnitude as that obtained for the Coulomb force. Such effect can arise from the residual strain of the cantilever and the asymmetry of the clamping (Figure 1). Note that the delay factor $Y(\omega)$ is essentially an imaginary number around the $\beta$-resonance, and therefore we expect a phase shift close to $\pi$ in that case. Instead, the instantaneous force acting on the $\alpha$-resonance is characterized by a $\pi/2$ shift at resonance. Therefore, our analysis was aided by the presence the photo-thermally induced $\beta$-resonance in Figs. 5e and 5f, providing a phase reference. Finally, in the hypothesis that the THz force exciting the $\alpha_1$ mode is of purely photo-thermal origin, the phase behavior is drastically different from Fig.5e (shown in Methods), unambiguously confirming the presence of an instantaneous Coulomb contribution to the optomechanical force.

In our device, both the instantaneous Coulomb force and the photo-thermal force arising from the bilayer structure can be used for fast THz detection at room temperature. Even though in the geometry of our experiments the Coulomb force has smaller amplitude than the force from the bilayer, it has the net advantage of not being limited by the thermal diffusion time constant $\tau$. This can be seen qualitatively from the data in Fig. 4b and 4c, where the responsivity of the $\alpha$-resonances catches up with that of $\beta$-resonances at higher frequencies. Indeed, using Eq. (2) and the expression of the Brownian noise we obtain that the expression of the *NEP* is provided by $NEP = \sqrt{8k_B T m_{eff} \Gamma_m} (\omega_{THz} d_{gap}^{eff} / Q_{THz})$ (16 nW/Hz$^{0.5}$ in the present case) and is independent from the frequency. Our resonator concept, which allows localizing the Eddy currents and heat generation into very small volumes, also allows reducing the thermal diffusion time, i.e. by reducing the cantilever dimensions[28]. Improved photon collection can be achieved through integrating our resonators with planar antennas[29], and *NEP* below the 100 pW/Hz$^{1/2}$ level becomes achievable. Both detection mechanisms are already suitable for room temperature applications with externally modulated sources and they allow detecting at frequencies that are much higher than those of conventional THz detectors. The planar geometry of our structure is also very convenient for large scale integration for imaging arrays[30], or for integrating the device on a single chip[31].

In conclusion we have introduced a metamaterial resonator where the strong sub-wavelength confinement allows converting electromagnetic energy into micromechanical motion at the nanoscale. Besides the dissipative photo-thermal forces, we have demonstrated a THz reaction (Coulomb) force that realizes the fundamental Hamiltonian of quantum opto-mechanics[8]. The reaction force can be optimized by increasing the quality factors of both THz and mechanical resonators. For instance, higher electromagnetic quality factors $Q_{THz}$ up to 100-200 are achievable in symmetric resonators[32]. When tested in vacuum (data not shown), the present cantilevers displayed mechanical quality factors on the order of $Q_m$ = 1000-3000. Furthermore our structure is based on semiconductor technology and operates in a frequency range where THz electronic transitions in quantum heterostructures can be achieved[3,33]. Therefore, beyond the detector application, our device concept could open new perspectives for fundamental studies of opto-mechanical and light-matter interactions in the THz range.

**Acknowledgment**

We acknowledge financial support from the ANR (Agence Nationale de la Recherche) and CGI (Commissariat à l'Investissement d'Avenir) through Labex SEAM "Capture" project (ANR 11 LABX 086, ANR 11 IDEX 05 02).We acknowledge useful discussions with Pierre Verlot (University of Lyon) on the NIR detection scheme. We also acknowledge technical help from Hua Li who fabricated the THz QCL.

**Author contribution**

**Methods**

**Sample fabrication protocol**

The composition of the wafers used to fabricate our structures is the following: a semi-insulating GaAs substrate, an epitaxially grown GaAs 320 nm buffer layer, a 1.6μm $Al_{0.8}Ga_{0.2}As$ sacrificial layer, and a 320 nm GaAs top layer. The structures are defined by e-beam lithography on PMMA 950 resist. After revealing the resist, a layer of 5 nm of platinum (Pt) and 150nm of gold (Au) are evaporated on the sample. Then, a lift-off is performed in order to define the metallic patterns that serve as an etch mask for dry etching in an inductively coupled plasma reactor. In this step the top GaAs and AlGaAs sacrificial layer are removed everywhere but under the features protected by the metal. The sample is then wet etched in a refrigerated (4°C) dilute HF 2.5% solution for 45s, which allows removing anisotropically the sacrificial layer of AlGaAs with a depth to lateral etching ratio of 1.3. To avoid the sticking of the cantilever to the substrate, the sample is put on a hot plate at 200°C for 60 s.

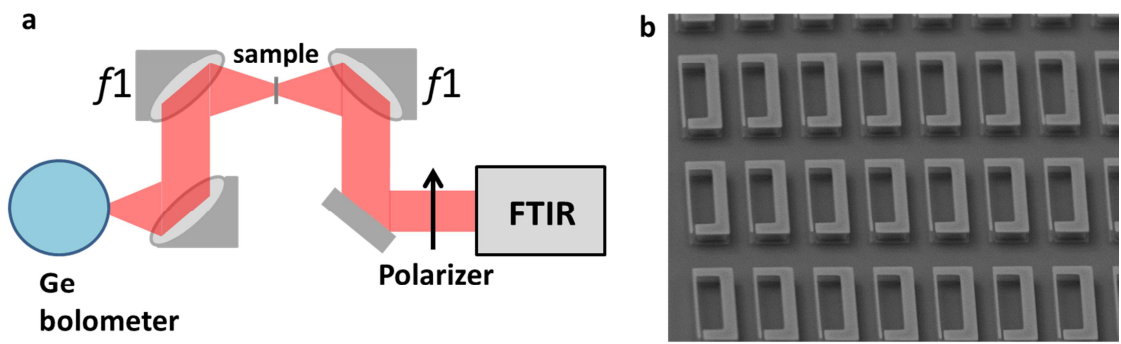

**Extended Data Figure 1**: **THz transmission spectroscopy a,** Experimental setup used for the transmission measurements on the SRR samples. **b**, Scanning electron microscope image of a dense array of identical resonators used for the transmission experiments.

**THz spectroscopy of SRR**

To infer the THz resonances of our split ring resonators, we have recorded transmission spectra with a Fourier Transform Interferometer (FTIR). The schematics of the experimental setup are shown in the Extended Data Fig.1a. Radiation from the Glow-bar source of the FTIR was focused and collected after passing through the structure using four f/1 parabolic mirrors, and finally detected by a cooled Ge bolometer. A polarizer positioned at the output of the FTIR allowed selecting the polarization of the incident beam. The whole set-up is placed in a purged environment in order to minimize water absorption. Measurements were performed on a dense array of resonators, as shown in Extended Data Fig. 1b, with 5μm spacing between the structures, in order to obtain clear signatures of the resonances in the transmission spectra[22, 23]. Indeed, the typical beam spotsize was of the order of 1mm², i.e. much larger than the typical cross section of a single resonator (4x10⁻⁴mm²). Spectra were normalized using a reference spectrum obtained from the transmission through the bare semiconductor substrate. A baseline correction was also applied.

In Fig. 2a, the SRR resonance appears as a transmission dip at 2.7 THz. The position of the resonance agrees well with the formula: $\omega_{THz}=\pi c/n_{eff}P$, where $c$ is the speed of light, $P$ =44.5 µm the perimeter of the resonator in Fig. 1, and $n_{eff}$ is an effective index. This formula assumes a $\lambda/2$ standing wave pattern with electric field antinodes at the plate sides of the SRR. Knowing the frequency $\omega_{THz}/2\pi$ =2.7 THz we obtain an effective index $n_{eff}$=1.27. This value compares well to the refractive index $(1+\alpha+(1-\alpha)\cdot n_{AlGaAs})/2$=1.21, where α is the ratio between the length of the cantilever (suspended part) and the perimeter $P$. The latter formula averages between the air and the refractive index of the AlGaAs layer below the resonator, with an index $n_{AlGaAs}$=3.18.

**Detection of the cantilever motion**

The cantilever motion is measured using an optical detection scheme based on a near infrared (NIR) $\lambda$=940nm laser diode. The laser beam is focused on the cantilever through a microscope objective. The fraction of the NIR light reflected by the cantilever is split in two spatially separated beams using a sharp edge blade. These beams are subsequently focused on a balanced photo-detection unit connected to a spectrum analyzer (SA). Generally speaking, the light intensity $I(y,z)$ scattered by the cantilever is a function of its in-plane ($y$) and out-of-plane ($z$) positions. Our optical detection setup provides the intensity fluctuations induced by the cantilever displacements with respect to the rest point of the latter: $\delta I = (\partial I/\partial y)y + (\partial I/\partial z)z = I_y y + I_z z$. Here $I_y$ and $I_z$ are real coefficients that depend on the position of the laser spot focused on the cantilever, and can be adjusted experimentally (Figures 5a and 5d). As a result the signal provided by the SA will be generally proportional to the linear combination $I_y S_{yy}(f)+I_z S_{zz}(f)$, where $S_{yy}$ and $S_{zz}$ are the spectral components of the Brownian noise power density of the cantilever displacements[8]. Furthermore, the total noise can be expressed as $\gamma S_{yy}(f)+ (1-\gamma^2)^{1/2}S_{zz}(f)+S_0(f)$, where $S_0$ is the noise floor of the photodetectors, and $\gamma=I_y/(I_y^2+I_z^2)^{1/2}$. Since we can experimentally set $\gamma$=0 or $\gamma$=1, and the noise floor is constant with frequency and much lower than the Brownian noise, we can use the analytical expressions for $S_{yy}(f)$ and $S_{zz}(f)$ in order to calibrate the SA readings. Finally, we note that the situation where $\gamma$=1/2 corresponds to the Boltzmann equipartition theorem.

In the set-up where the SRR was coupled with a modulated THz QCL (Fig. 4), the output of the balanced detection was either sent to a spectrum analyzer, or to a lock-in amplifier referenced to the QCL modulation frequency $f_{mod}$.

**Mechanical characteristics of the cantilever**

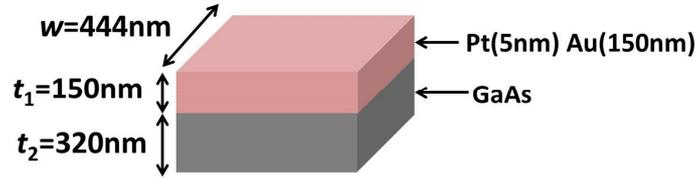

**Extended Data Figure 2:** Schematics of the cross section of the cantilever part

The cross section and material composition of our cantilever is indicated in the Extended Data Figure 2. As shown in Fig. 1 the length $L$=15.7µm of the cantilever is measured from the free end to the mid-point of the etched pedestal. Regarding the cantilever as a homogeneous object, the resonance of the first in-plane flexural mode is described by the equation[34] $f_1$=0.162$(w/L^2)(Y/\rho)^{1/2}$. Here $Y$ is the Young's modulus and $\rho$ the volume density of the cantilever. Neglecting the 5 nm Pt adhesive layer indicated in Extended Data Fig. 2, the main constituents of the cantilever are Gold (Au) and GaAs, with the following characteristics:

$$Au\begin{cases} Y_{Au} = 79 GPa \\ \rho_{Au} = 19.3 g/cm^3 \end{cases}, \quad GaAs\begin{cases} Y_{GaAs} = 85.5 GPa \\ \rho_{GaAs} = 5.3 g/cm^3 \end{cases}$$

As the Gold and GaAs have very similar Young's moduli, we take the average $Y$=0.5$(Y_{Au}+Y_{GaAs})$=82.25 GPa. Similarly, we use a mean density averaged over the thickness of the different materials: $\rho = (\rho_{Au}t_1+\rho_{GaAs}t_2)/(t_1+t_2)$ = 9.8 g/cm$^3$. Using these values, we obtain a numerical estimate $f_{\alpha 1}$=845 kHz, which compares very well with the experimental value 860 kHz, considering the fact that dimension are known with 10% uncertainty. Similarly, the frequency of the out-of-plane mode is provided by the formula $f_{\beta 1}$=$f_{\alpha 1}$ $(t_1+t_2)/w$=910 kHz, and it is in excellent accordance with the experimental value. This justifies our approach of using average density and Young modulus for the cantilever. The frequencies of the higher order modes are obtained from the expression $f_{(\alpha/\beta)n}$=2.81$f_{(\alpha/\beta)1}(n-0.5)^2$, also in excellent agreement with the experimental values $f_{\alpha 2}$=5.4 MHz and $f_{\beta 2}$=5.7 MHz, meaning that the cantilever geometry is known with high precision .

The effective mass $m_{eff}$ of the cantilever is obtained according to the formula[34] $m_{eff}$ = (33/140)$\rho wt$ = 8.5 pg. This mass corresponds to the endpoint movement of the cantilever, and we checked numerically that it is constant for all the flexural modes considered in this work. The frequency can thus be re-expressed as $f_1$ = $(1/2\pi)(k/m_{eff})^{1/2}$ where $k$ = 0.25 N/m is the effective spring constant (similar for both fundamental modes). Using the parameters $k$ and $m_{eff}$ defined above, we can treat the cantilever as a one dimensional harmonic oscillator. Within this picture, we can also express the noise spectral density of the Brownian motion of the cantilever $S_{yy}(f)$ as a function of frequency[8], for instance for the $\alpha_1$ mode we obtain:

$$S_{yy}(f) = \frac{2k_B T \Gamma_m}{m_{eff}(2\pi)^3} \frac{1}{(f^2 - f_{\alpha_1}^2)^2 + (f\Gamma_m)^2}$$

A similar expression holds for $S_{zz}(f)$ and all other modes. Fitting the experimental spectra of the room temperature Brownian motion with the above equation allows, on one side, inferring the broadening parameter $\Gamma_m$ for each mode, and, on the other side, calibrating the signal from the spectrum analyzer, as described above and in the main text.

### THz optomechanical coupling and effective capacitance

The coupling between the SRR and the cantilever movement can be described in the picture of an equivalent LC circuit[12,14]. The electric field in the gap shown in Fig. 1 is described by an equivalent capacitor $C(y)$ which is a function of the cantilever's displacement $y$. Therefore, the dynamical variables of the coupled system are $y$, the displacement of the cantilever tip, and $q$, the charge induced on the capacitor plates. Then the system's Lagrangian $\mathscr{L}$ is written as:

$$\mathscr{L} = \frac{1}{2}m_{eff}\dot{y}^2 + \frac{1}{2}L\dot{q}^2 - \frac{q^2}{2C(y)} - \frac{1}{2}m_{eff}\omega_m^2 y^2$$

The cantilever movement can then be provided, for instance, by the Lagrange equation for the position:

$$m_{eff}\frac{d^2 y}{dt^2} + \frac{m_{eff}\omega_m}{Q_m}\frac{dy}{dt} + m_{eff}\omega_m^2 y = \frac{d}{dy}\left(\frac{q^2}{2C(y)}\right)$$

Usually, the displacement of the cantilever $y$ is small compared to the gap $d_{gap}$, therefore the spatial derivative in the above equation can be approximated at zeroth order with its value at $y=d_{gap}$. Using the expression for the total electrical energy stored in the resonator $W_{THz} = q^2/2C(y)$, and introducing an effective gap through the formula $d_{gap}^{eff} = ((d\ln C(y)/dy)|_{y=d_{gap}})^{-1}$ we arrive at the results stated in the main text (Eq.(1)). The parameter $d^{eff}_{gap}$ defined above allows to evaluate the magnitude of the Coulomb force for an arbitrary geometry. However, its exact analytical evaluation is difficult, due to the fringing fields of the capacitance. To determine $d^{eff}_{gap}$, we performed 2D quasi-static numerical simulations of the cross section of the structure in a plane perpendicular to the cantilever, using a finite element method. We evaluated the electric energy of the system $W_e$, considering that the metallic part of the cantilever has a potential of 1V, while the opposite metallic side is grounded. Performing such simulations for various positions $y$ of the cantilever we obtained the effective gap $d_{gap}^{eff} = ((d\ln W_e(y)/dy)|_{y=d_{gap}})^{-1}$=800nm. The electric field maps obtained in that case were very similar to the full electromagnetic simulation of the THz resonance mentioned in the main text.

Such simulations, provided a negligible Coulomb force for cantilever displacements along the z-direction .

The fact that the effective gap (800nm) is larger than the physical gap (308nm, see Fig.1) is understandable, since the fringing fields which "run away" from the structure are expected to be less sensitive to the cantilever displacement compared to the field confined in the gap . Note also that owing to propagation effects in the actual resonator the charge density develops along the whole cantilever length. Here we have considered only the charges in the vicinity of the gap, since the effects of the other parts are expected to be small as the distance between the positive and negative charges is much larger than $d_{gap}$.

**THz QC Laser characteristics and intensity modulation**

The QCL used in this experiment has an active region similar to the one described in Ref. **25**. It was processed in a 3 mm-long single metal ridge waveguide. The QCL threshold current is $I_{th}$=1.0A, and the output power is of the order of a few mW at a heat sink temperature of 20 K. Using a calibrated THz power meter (Ophir 3A-P-THz ROHS) we rerecorded a maximum emitted power of 4.7 mW. In Extended Data Fig. 3 we show the spectrum of the QCL, recorded at V= 48.5V superimposed to the transmission spectrum of the SRR array from Extended Data Fig. 2b. The QCL has a central emission frequency of 2.6 THz, slightly detuned from the SRR resonance at 2.7 THz, still within the SRR resonant bandwidth.

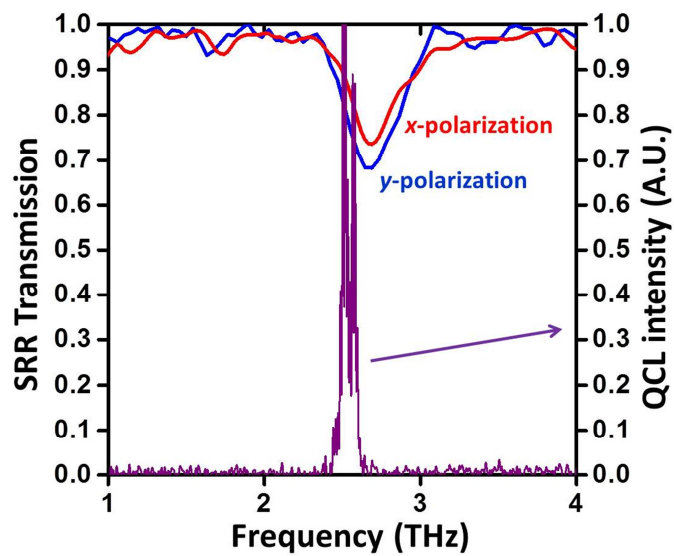

**Extended Data Figure 3**: **QCL and SRR spectra.** QCL emission spectrum superimposed to the transmission spectrum of the SRR arrays from Fig. 2a. "x/y-polarization" refers to transmission measurements with an electric field vector along the x/y axis as defined in Fig.1. In our experiments, the laser electric field was always along the *y*-direction.

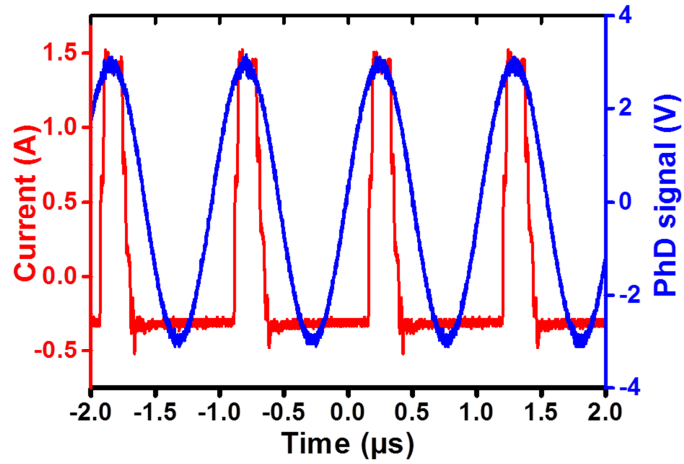

**Extended Data Figure 4: Time traces representative of the incident THz intensity and the corresponding mechanical response of the cantilever.** In red: QCL driving current. In blue: signal form the balanced photo detection unit, recorded on an electronic oscilloscope.

To modulate the driving THz force on the cantilever, the QCL's current was modulated from threshold to the maximum output power with a square wave with a duty cycle of 16.5%. In Extended Data Fig. 4 we show the time trace of the driving current (red curve) measured with an electronic oscilloscope. When the modulation frequency was matched to the mechanical frequency of the cantilever, the latter showed harmonic oscillations, as shown by the photocurrent from the balanced detection unit (blue curve). This means that the mechanical resonator responds to the first Fourier harmonic of the square wave intensity modulation. The latter has an amplitude that is $\sin(0.16\pi)/\pi$ of the square wave amplitude, and was used for the experimental estimation of the detector *NEP* described in the main text.

To verify the resonant interaction between the THz source and the SRR, we also performed the detection experiment described in the main text using another QCL, which was emitting at 2.0 THz and was therefore detuned from the SRR resonance. The 2.0 THz QCL delivered a maximum power of 1mW and was operated in the same conditions as the 2.6 THz QCL: pulsed mode with a 16.5%, duty cycle, and a modulation frequency matched to the cantilever mechanical resonance. As expected, using this laser we could not excite coherently the cantilever. Indeed, the QCL frequency being too far away from the SRR resonance, charge and current oscillations leading to a mechanical force on the cantilever could not be excited.

**Modelling of the dynamic photo-thermal force induced by the Eddy currents**
The THz Eddy currents shown in Figure 2c induce an inhomogeneous heating described by a temperature profile $\Delta T(x)$, with the coordinate system Oxyz defined in Fig.1 (Exceptionally, in this part z is a local coordinate and the cantilever displacement is noted $\delta z$). The first step of our analysis is to determine the total elastic energy stored in the cantilever that corresponds to this heating effect. The elastic energy density can be expressed as $u(x,z) = -Y_i\gamma_i \Delta T(x)\varepsilon(x,z)$[35] with $i=1$ or 2 depending on whether $z$ is in the Au or GaAs part, $Y_i$ are the corresponding Young's moduli, $\gamma_i$ the thermal expansion coefficients, and $\varepsilon(x,z)$ the deformation. Taking the deformation of a bended cantilever as $\varepsilon(x,z) = (z-z_0)/R(x)$, where $z_0 = (t_1+t_2)/2$ and $R(x)$ is the local curvature radius, and by integrating over the cantilevers thickness we obtain for the total thermoelastic energy:

$$U = -wY \frac{(\gamma_1 - \gamma_2)t_1 t_2}{2} \int_0^L \frac{d^2 \delta z}{dx^2} \Delta T(x) dx$$

Here $Y$ is the average Young modulus and we replaced the curvature with its approximate value $1/R(x) = d^2\delta z/dx^2$, the function $\delta z(x)$ describing the shape of the bended cantilever.

Actually, the thermal profile $\Delta T(x)$ depends not only on $x$ but $z$ as well. In the following, we determine $\Delta T(x,z)$ and we define an average over the cantilever thickness $\Delta T(x) = \frac{1}{(t_1+t_2)} \int_0^{t_1+t_2} \Delta T(x,z) dz$ that can be used directly with the expression for the thermoelastic energy.

To determine the full temperature profile we solve the Fourier heat equation in a steady state with a periodic heat source excitation at the top of the Au layer. This is justified as at THz frequencies heat is generated by the Eddy currents within the skin depth of the metal. We first consider a temperature profile $\Delta T(x,z)\exp(i\omega t)$ that satisfies the heat equation: $i\omega \Delta T / D = \partial^2 \Delta T / \partial x^2 + \partial^2 \Delta T / \partial z^2$, with $D = (\lambda_1 t_1 + \lambda_2 t_2)/(c_1 \rho_1 t_1 + c_2 \rho_2 t_2)$ the thermal diffusion coefficient, $\lambda_i$ the thermal conductivities and $c_i$ the specific heat capacitances. The boundary conditions applied are: zero heat flow on the lateral surfaces of the cantilever and at its free end, except for the top Au surface. Since we consider only the dynamic part of the temperature profile we take $\Delta T = 0$ at the clamping point. The boundary condition on the Au surface is written: $-\lambda_1 \partial \Delta T / \partial z = P_{THz}(x)/(L\Sigma)$ with $\Sigma = wL$ the top area of the cantilever. The function $P_{THz}(x)$ is normalized such that $\int_0^L P(x)dx$ is the total power $P_{THz}$ dissipated by the Eddy currents, which is basically the power coupled by the SRR. The solution of the heat equation with these boundary conditions is expressed by a series of cosine and hyperbolic functions. By averaging $\Delta T(x,z)$ over the cantilever thickness we obtain:

$$\Delta T(x) = \frac{1}{(t_1+t_2)} \int_0^{t_1+t_2} \Delta T(x,z) dz = \frac{-P_{tot} 4L^2}{\lambda_1 \pi^2 \Sigma (t_1+t_2)} \sum_{n=0}^{\infty} \frac{p_n \sin\left(\frac{\pi x}{L}(n+0.5)\right)}{i\omega \tau_0 + (2n+1)^2}$$

Here $p_n = \int_0^L P(x)\sin\left(\frac{\pi x}{L}(n+0.5)\right)dx / P_{THz}$ are series of dimensionless coefficients describing the projection of the heat source on the spatial harmonics of the temperature profile. First, we include this expression in the formula of the elastic energy above. Second, we define a dynamical profile if the cantilever vibrational mode $\delta z(x) = z_{max}Q_\beta(x)$, where $z_{max}$ is the amplitude of the displacement of the free end of the cantilever, and $Q_\beta(x)$ is a dimensionless eigen-function which is solution of the one-dimensional equation for elastic waves[35], normalized such that $Q_\beta(y) = 1$ at the free end. The total thermoelastic energy thus becomes:

$$U = \frac{2Y(\gamma_1 - \gamma_2)t_1 t_2}{\pi^2 \lambda_1 (t_1 + t_2)} P_{THz} z_{max} Y(\omega)$$

The function $Y(\omega)$ is defined in Eq.(3) of the main text with the coefficients $A_n = p_n r_n$. Here $r_n$ is the dimensionless projection integral $r_n = L\int_0^L \frac{d^2 Q_\beta}{dx^2}\sin\left(\frac{\pi x}{L}(n+0.5)\right)dx$. The expression of the effective photothermal force $f_{ph}$ acting on the cantilever end is finally given by:

$$f_{ph} = \frac{\partial U}{\partial z_{max}} = \frac{2Y(\gamma_1 - \gamma_2)t_1 t_2}{\pi^2 \lambda_1 (t_1 + t_2)} P_{THz} Y(\omega)$$

Since $f_{ph}$ is proportional to the absorbed power, we can define a force $f_{ph}^0$ per unit absorbed THz power as in Eq.(3). For the models presented in this paper the absorbed power (Joule heating) is modelled with the spatial profile $sin^2(x\pi/L)$, which fits the current profiles provided by 3D electromagnetic simulations illustrated in Fig. 2c.

**Modelling of the residual thermal strain for the in-plane modes**

The residual thermal strain for the in-plane mode is modelled as described in Ref.[27], that is as a force retarded by the thermal diffusion time $\tau$. The frequency dependent expression of this force is then $C_{ph}/(1+i\omega\tau)$[27]. This is very similar to the photo-thermal force associated to the bilayer; indeed the function $Y(\omega)$ describes the retardation effects owing to the finite heat diffusion time across the cantilever length. However, no analytical expression for $C_{ph}$ is known, and the latter is therefore treated as a fit parameter.

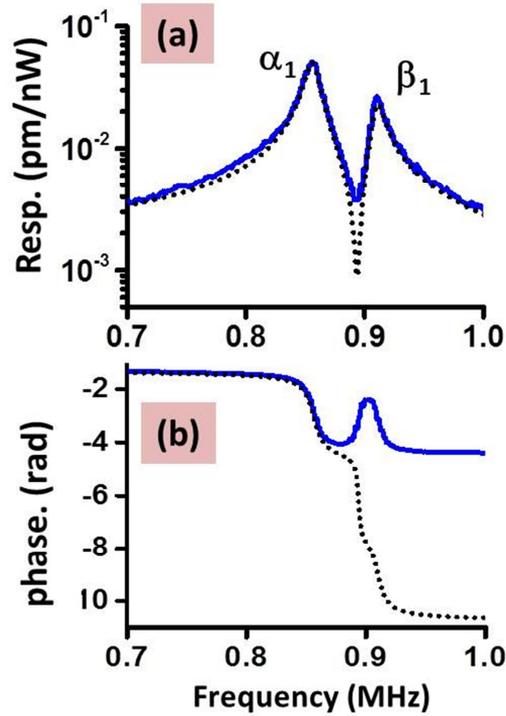

**Extended Data Figure 5:** Fit of the amplitude (a) and phase (b) of the same data as in 5e and 5f, in the main text, but assuming purely photo-thermal forces for both $\alpha_1$ and $\beta_1$ modes (dotted curve).

The modelling of the data in the hypothesis that only photo-thermal forces are present is shown in Extended Data Fig. 5. We first adjust $C_{ph}$ so that the amplitudes of both the $\alpha_1$ and $\beta_1$ modes are recovered (Extended Data Fig. 5a). The resulting phase, shown in Extended Data Fig. 5b, is very different from the measured one.

**Collection efficiency of the Si lens**

The numerical aperture of our lens was 0.60, providing a FWHM equal to 23µm for the Airy spot inside the GaAs substrate of refractive index 3.5, for a THz QCL laser wavelength $\lambda$=115µm (2.6THz). The corresponding area is 1640µm². The collection area of the SRR is estimated from the reflectivity data in Extended Data Figure 3. The array unit cell in that case has an area of 308µm², and the resonance contrast is 0.35. This provides a collection area of 108µm² per resonator. The coupling efficiency is then estimated at 108*0.7/1640=4.6%, where the factor 0.7 is the air to Si transmission coefficient. Note that a larger coupling efficiency is expected for a single resonator as compared to a dense array[22].